\documentclass[aps, prd,11pt,notitlepage,nofootinbib,superscriptaddress,showkeys]{revtex4-1}
\usepackage{amstext,amsmath,amssymb,amsfonts,bbm, amsthm}
\usepackage[latin1]{inputenc}
\usepackage{fancyhdr}
\usepackage{graphicx}
\usepackage{hyperref}
\usepackage{tocvsec2}

\usepackage{color}

 \usepackage[usenames,dvipsnames]{pstricks}
 \usepackage{epsfig}
 \usepackage{pst-grad} % For gradients
 \usepackage{pst-plot} % For axes

\def\beq{\begin{equation}}
\def\be{\begin{equation}}
\def\ee{\end{equation}}
\def\bes{\begin{eqnarray}}
\def\ees{\end{eqnarray}}

\def\f{\frac}

\begin{document}
%%%%%%%%%%%%%%%%%%%%%%%%%%%%%%%%%%%%%%%%%%%%%%%%%%%

\title{The two faces of Hawking radiation\footnote{Honorable Mention in the Gravity Research Foundation 2013 Essay Competition}}

\author{{Matteo Smerlak}}\email{smerlak@aei.mpg.de}
\affiliation{Max-Planck-Institut f\"ur Gravitationsphysik, Am M\"uhlenberg 1, D-14476 Golm, Germany}

\date{\small\today}

%%%%%%%%%%%%%%%%%%%%%%%%%%%%%%%%%%%%%
\begin{abstract}\noindent
What happens when Alice falls into a black hole? In spite of recent challenges by Almheiri \emph{et al.}---the ``''firewall'' hypothesis---, the consensus on this question tends to remain ``nothing special''. Here I argue that something rather special can happen near the horizon, already at the semiclassical level: besides the standard Hawking outgoing modes, Alice can records a quasi-thermal spectrum of \emph{ingoing} modes, whose temperature and intensity diverges as Alice's Killing energy $E$ goes to zero. I suggest that this effect can be thought of in terms a \emph{horizon-infinity duality}, which relates the perception of near-horizon and asymptotic geodesic observers---the two faces of Hawking radiation.
\end{abstract}
%%%%%%%%%%%%%%%%%%%%%%%%%%%%%%%%%%%%%%
\maketitle

\section{Introduction: Alice falls into a black hole}

Everybody knows \cite{Hawking:1988wo} that a black hole---the final state of gravitational collapse \cite{Oppenheimer:1939ud}---emits a stationary flux of massless particles \cite{S1975}. From the perspective of \emph{asymptotic} inertial observers, this Hawking flux is indistinguishable from the thermal radiation emitted by a black body at temperature $T_{H}=\kappa/2\pi$ (in Planck units), where $\kappa=(4M)^{-1}$ is the surface gravity of the hole. For macroscopic black holes (with $M\gtrsim M_{\odot}$), this value is exceedingly small---indeed much smaller than the CMB temperature; it is therefore unlikely that Hawking radiation will ever be measured directly (at least in the astrophysical context.\footnote{See however \cite{Barcelo2011} for a review of promising experimental attempts based on Unruh's \cite{Unruh1981} ``analogue gravity'' proposal.})

But what if one could get very close to a hole? What would Hawking radiation look like \emph{there}? Unruh asked this question in his celebrated ``Notes on black hole evaporation'' \cite{Unruh1976}, where he showed that the correct answer depends on the observer's state of motion: the temperature measured by a particle detector close to the horizon is $T_{U}=a/2\pi$, where $a$ is the detector's acceleration. Consequently, according to Unruh, ``a geodesic detector near the horizon will not see the Hawking flux of particles'' \cite{Unruh1976}. Because the near-horizon of a sufficiently massive black hole is almost flat, and the quantum vacuum is as regular there as anywhere else, it is indeed natural to assume that ``nothing special'' happens to geodesic observers at the horizon. 

Yet, this consensual view has been recently challenged on various grounds. For Helfer \cite{HELFER2004}, taking interactions into account ``completely alters the picture'' drawn by Hawking and Unruh: ultra-high-energy vacuum fluctuations must couple to quantum gravity, and thus quantum field theory must break down in the vicinity of a black hole. For Almheiri \emph{et al.} \cite{Almheiri2012} and Braunstein \emph{et al.} \cite{Braunstein:2013vr}, the devil is in the information conservation principle: for Hawking radiation to be in a pure state, something ``dramatic'' must happen at the horizon. Pictorially, if Alice is entangled with Bob but---being somewhat braver than him---decides to cross the horizon and plunge into the hole, then quantum mechanics will ``burn her up''. 

These new arguments are both exciting and challenging, but let's face it: they are mere speculations. To this day, neither the physics of trans-Planckian fluctuations nor the fate of unitarity in quantum gravity are properly understood; drawing dramatic consequences from them is bold, but risky. What \emph{is} well understood, on the other hand, is quantum field theory in curved spacetimes---the very setup used by Hawking to make his prediction. What does the old semi-classical framework have to say about Alice falling into a black hole?

\section{Believing impossible things before breakfast: will Alice burn up?}

\begin{figure}[t]
\begin{center}
\vspace{.6cm}
\scalebox{1.1} % Change this value to rescale the drawing.
{
\begin{pspicture}(0,-2.958047)(5.50291,2.958047)
\psline[linewidth=0.04cm](0.6210156,2.5596485)(0.6210156,-2.4403515)
\psline[linewidth=0.04cm](0.6210156,-2.4403515)(3.8210156,0.75964844)
\psline[linewidth=0.04cm](3.8210156,0.75964844)(2.2210157,2.5596485)
\psline[linewidth=0.04cm,linestyle=dashed,dash=0.16cm 0.16cm](0.6210156,2.5596485)(2.2210157,2.5596485)
\psdots[dotsize=0.12](1.8210156,1.1596484)
\psdots[dotsize=0.12](1.8210156,1.1596484)
\psline[linewidth=0.02cm](1.8210156,1.1596484)(3.0210156,-0.040351562)
\psline[linewidth=0.02cm](1.8210156,1.1596484)(0.6210156,-0.040351562)
\psline[linewidth=0.02cm](0.6210156,-0.040351562)(1.8210156,-1.2403516)
\usefont{T1}{ptm}{m}{n}
\rput(1.9924707,1.4246484){$x$}
\usefont{T1}{ptm}{m}{n}
\rput(3.6724708,-0.115351565){$v_{+}(x)$}
\usefont{T1}{ptm}{m}{n}
\rput(2.3524706,-1.4953516){$v_{-}(x)$}
\psline[linewidth=0.04cm,linestyle=dotted,dotsep=0.16cm](2.2210157,2.5596485)(0.64101565,1.0396484)
\usefont{T1}{ptm}{m}{n}
\rput(0.7824707,-2.7353516){$i_-$}
\usefont{T1}{ptm}{m}{n}
\rput(2.4824708,2.7646484){$i_+$}
\usefont{T1}{ptm}{m}{n}
\rput(4.1724706,0.76464844){$i_0$}
\usefont{T1}{ptm}{m}{n}
\rput(2.9824708,-0.79535156){$\mathcal{J}^{-}$}
\usefont{T1}{ptm}{m}{n}
\rput(3.5824707,2.0046484){$\mathcal{J}^{+}$}
\psline[linewidth=0.03cm,arrowsize=0.05291667cm 2.0,arrowlength=1.4,arrowinset=0.4]{->}(3.9610157,-1.2203516)(5.0010157,-0.14035156)
\usefont{T1}{ptm}{m}{n}
\rput(5.1924706,-0.39535156){$v$}
\usefont{T1}{ptm}{m}{n}
\rput(1.2524707,2.1646485){$\mathcal{H}^+$}
\end{pspicture} 
}
\caption{Penrose diagram of gravitational collapse and definition of the ``eikonal coordinates'' $(v_{+},v_{-})$.}
\label{penrosediag}
\end{center}
\end{figure}
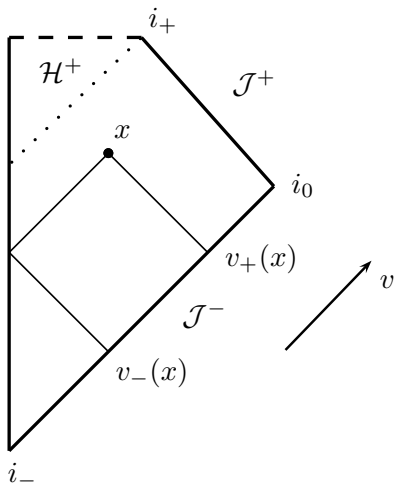

Consider the simplest possible model of gravitational collapse---an ingoing null shell with ADM mass $M$ coupled to a free, massless scalar field in $1+1$ dimensions---and the simplest possible model of Alice---an Unruh-DeWitt \cite{Unruh1976,DeWitt1979} detector with energy gap (frequency) $\Omega$ moving along a radial geodesic $\gamma(\tau)$ with conserved energy (per unit rest mass) $E$. In this setup, the answer to ``what does Alice see at time $\tau$?'' is coded in the response function 

\be\label{response}
\mathcal{R}(\tau,\Omega)=2\,\textrm{Re}\,\int_{-\infty}^{\infty}du\,\chi_{\tau}(u)\int_{0}^{\infty}ds\, \chi_{\tau}(u-s)\,e^{-i\Omega s}\,G\big(\gamma(u),\gamma(u-s)\big).
\ee
Here $G(x,y)$ is the two-point function of the field and $\chi_{\tau}(u)$ a smooth, non-negative ``window function'', such that $\int_{-\infty}^{\infty}\chi_{\tau}(u)du=1$ and $\chi_{\tau}(u)\simeq0$ for $u\geq\tau$. To compute \eqref{response} for a given radial geodesic $\gamma(\tau)$, only one ingredient from quantum field theory is needed: the two-point function $G(x,y)$ of the field in the in-vacuum.    

The in-vacuum state is best described in terms of what may be called ``eikonal coordinates'' $(v_{+},v_{-})$. These are globally defined, null coordinates, defined for a given spacetime point $x$ by tracing back to past null infinity $\mathcal{J}^{-}$ the two radial null rays which meet at $x$; the coordinates $v_{\pm}(x)$ are the Eddington-Finkelstein coordinates $v=t+r^{*}$ of these rays on $\mathcal{J}^{-}$, see Fig. \ref{penrosediag}. In these coordinates, the two-point function $G(x,y)$ in the in-vacuum takes the simple form \cite{Birrell1982,Helfer2003}
\begin{eqnarray}
G(x,y)&&\propto\log\Big((v_{+}(x)-v_{+}(y)-i0)(v_{-}(x)-v_{-}(y)-i0)\Big).
\end{eqnarray}
Notice that, by virtue of the additivity of the $\log$, in $1+1$ dimensions Alice's detector couples independently to $v_{+}$ (the \emph{incoming} vacuum modes) and $v_{-}$ (the \emph{outgoing} vacuum modes). This property does not hold in higher dimensions, or with massive fields, but for the argument made in this essay, it is key.  

Let us now ask: when will either of the two contributions $\mathcal{R}_{\pm}$ to the response function \eqref{response}---corresponding to the incoming ($\mathcal{R}_{+}$) and outgoing ($\mathcal{R}_{-}$) modes respectively---be thermal? There is a simple criterion for this \cite{Barcelo2011a,Barbado2011,Smerlak:2013vb}: it suffices that, along the trajectory $\gamma(\tau)$, the quantity (dot denotes $d/d\tau$)
\be\label{quasi}
T_{\pm}=\f{1}{2\pi}\Big|\f{\ddot{v}_{\pm}}{\dot{v}_{\pm}}\Big|
\ee
is approximately constant, in which case $\mathcal{R}_{\pm}$ is a thermal spectrum at temperature $T_{\pm}$. Here, by ``approximately constant'' I mean $\vert\dot{T}_{\pm}/T_{\pm}^{2}\vert\ll1$. When this condition does not hold, a weaker form ``thermality'' of $\mathcal{R}_{\pm}$ still holds in the ultraviolet limit, namely $\mathcal{R}_{\pm}\propto e^{-\Omega/T_{\pm}}$ for $\vert\Omega\vert\gg T_{\pm}$. 

In the case of asymptotic inertial trajectories with $E=1$, as in Hawking's original work, one checks that $T_{+}^{\textrm{asymp}}=0$ and $T_{-}^{\textrm{asymp}}=T_{H}$. (This is consistent the lore that Hawking radiation consists of those modes which have bounced off the center of the shell and escaped from it just before it forms a horizon at $r=2M$). More generally, if $E>1$, viz. for inertial observers moving towards the hole at velocity $\dot{r}=\sqrt{E^{2}-1}$, one computes $T^{\textrm{asymp}}_{+}=0$ and
\be
T^{\textrm{asymp}}_{-}(E)=T_{H}\Big(E+\sqrt{E^{2}-1}\Big).
\ee
The $E$-dependent correction is the Doppler factor for an observer moving relative to the a static source located on the center of the black hole.

\begin{figure}[t]
\vspace{.6cm}
\begin{center}
\hfill
\includegraphics[scale=.9]{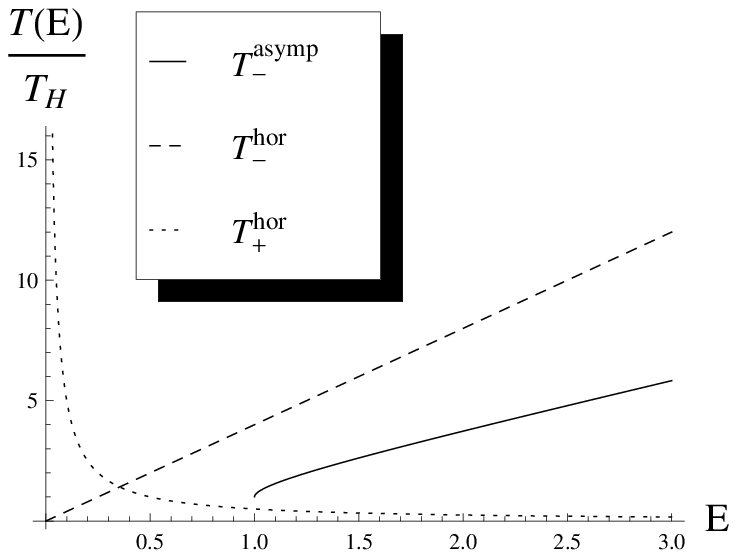}
%\hfill
%\begin{minipage}
\includegraphics[scale=.95]{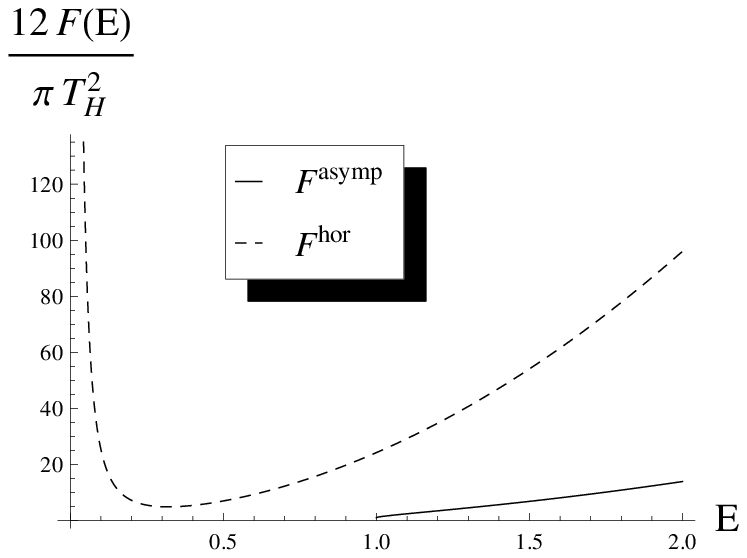}
%\end{minipage}
\caption{The ``temperature'' of outgoing and incoming modes (left) and outgoing flux (right) along a radial geodesic with energy $E$, both in the asymptotic limit (``asymp'') and at the horizon (``hor'').}
\end{center}
\end{figure}

But what are the values of \eqref{quasi} when the radial geodesic $\gamma(\tau)$ crosses the horizon---that is, when Alice jumps into the hole? An explicit computation detailed in \cite{Smerlak:2013vb} shows that, when she crosses the Schwarzschild horizon,
\be
T^{\textrm{hor}}_{-}(E)=4ET_{H}\qquad\textrm{and}\qquad T^{\textrm{hor}}_{+}(E)=\f{T_{H}}{2E}.
\ee
Thus, if Alice enters the hole with sufficiently low radial velocity, in the sense that $E\ll1$, (with $E\rightarrow0$ corresponding to the limit case where she is dropped from the horizon with zero initial velocity), she perceives the \emph{incoming} modes at \emph{high temperature}! This divergence is confirmed by the computation of the (outgoing) \emph{flux} $\mathcal{F}(E)=-\langle T_{ab}\rangle u^{a}n^{b}$, where $u^{a}$ is the $4$-velocity of Alice and $n^{b}$ is a unit vector normal to $u^{a}$: at horizon-crossing, this flux reads \cite{Smerlak:2013vb}
\be
\mathcal{F}^{\textrm{hor}}(E)=\pi T_{H}^{2}\Big(2E^{2}+\f{1}{48E^2}\Big). 
\ee
Thus, the intensity of Hawking radiation perceived by freely-falling observers near the horizon can in fact be arbitrarily high. Observer moreover that the velocity-dependence of $T^{\textrm{hor}}_{\pm}(E)$ and $\mathcal{F}^{\textrm{hor}}(E)$ is \emph{not} of Doppler-type---at least not with respect to a single source inside the black hole.

\section{Looking through the looking glass: the horizon-infinity duality}

What is going on? In short---the Hawking effect, in reverse. 

The mechanism for the conversion of vacuum fluctuation into thermal radiation in the standard Hawking effect (observed from infinity) is well known: it is the \emph{exponential redshift} experienced by outgoing modes on their way from the to-be horizon to infinity. This is evident from our criterion \eqref{quasi}: if $T_{-}\propto \ddot{v}_{-}/\dot{v}_{-}$ is constant, then by integration $v_{-}(\tau)$ is an exponential function of $\tau$. 

Now, for a near-horizon observer, it is clear that outgoing modes have had no chance to experience such exponential redshift. Indeed, this is precisely what Unruh understood in \cite{Unruh1976}, and what led him to conclude that a near-horizon observer will not see the Hawking flux of radiation. But one must not forget that the field also contains incoming modes! These modes, coming from infinity all the way down to the horizon, \emph{do experience a very large frequency shift}---an exponential blueshift. This is why $T_{+}^{\textrm{hor}}$ is not zero on the horizon, and this is also why $T_{\pm}^{\textrm{hor}}$ have a non-Dopplerian dependence on Alice's velocity with respect to the hole: in the $E\rightarrow0$ limit, the ``source'' is infinity---not the hole.\footnote{Note however that the the ingoing modes which dominate the $E\rightarrow0$ limit contributed \emph{positively} to the outgoing flux $\mathcal{F}(E)$.}

Already at the purely classical level, it is well-known that in her free-fall approach to the black hole horizon, Alice will feel the sky ``falling down'' on her head: when looking in the direction transverse to her radial motion, she will see an image of the entire sky compressed in a thin layer just above horizon.\footnote{Riazuelo's animations, available at \url{http://www2.iap.fr/users/riazuelo/bh/index.html}, give a striking impression of this optical effect.} The near-horizon Hawking effect discussed in this essay is reminiscent of this (apparent) inversion effect: seen from a radially infalling geodesic, asymptotic infinity and the horizon itself appear \emph{dual} to one another, with the outgoing field modes mapped to the incomes modes and the large $E$ regime to the small $E$ regime. Explicitly,
\be
T_{+}^{\textrm{hor}}(E/2)\sim T_{-}^{\textrm{asymp}}(1/E)\quad\textrm{as}\quad E\rightarrow0.
\ee
This identity is, I believe, a rather remarkable feature of Hawking radiation.
  
One may object that this ``horizon-infinity duality'' is imperfect. First, $T_{-}^{\textrm{hor}}\neq0$, while $T_{+}^{\textrm{asymp}}=0$; second, the thermality of incoming modes is not perfect (because $\vert\dot{T}_{+}^{\textrm{hor}}/(T_{+}^{\textrm{hor}})^{2}\vert\neq0$, but rather $\simeq1$, hence thermality holds for high-frequency detectors only); and third, because the $E$-dependence of $T_{-}^{\textrm{asymp}}(E)$ and of $T_{+}^{\textrm{hor}}(E)$ holds only for $E\ll1$ and $E\gg1$. One reason for this asymmetry is clear: the local spacetime geometry at infinity and on the horizon are not the same---the former is flat, the latter is not. Suppose however that instead of the Schwarzschild spacetime, we considered an artificial black hole geometry in which both infinity \emph{and} the horizon are flat (it is easy to design such a metric and repeat all computations using it instead of Schwarzschild\cite{Smerlak:2013vb}). Then the local geometries at infinity and on the horizon would be the same, and computing $T_{\pm}$ in both limits would lift the first caveat. I do not know whether the other two caveats are of importance.   

\section{Conclusion}

After many years of work on the Hawking phenomenon, the nature of Hawking radiation at the horizon---the fact that it appears to be emitted from \emph{outside} the hole, and can have \emph{high} temperature---still surprises us today. Not only is it not always true that ``nothing special'' happens on the horizon, but one needs not go beyond the semiclassical approximation to find some drama near the horizon. 

I have attempted to account for this effect by what I called the ``horizon-infinity duality'': the idea that the Hawking effect at the horizon is a reversed image of the standard Hawking effect, where the role of ingoing and outgoing modes, and large-energy and small-energy geodesics, are reversed. The emerging picture is intriguing: for Alice dropped just above the horizon, it's not the black hole that looks like a blackbody---it's the sky. 

\section*{Acknowledgments}

I thank Thanu Padmanabhan for drawing my attention to the fate of infalling detectors with regards to Hawking radiation, Suprit Singh for the collaboration leading to \cite{Smerlak:2013vb}, and Carlo Rovelli and Antonin Coutant for several useful discussions on black hole physics. I also note that the particular result $T_{-}^{\textrm{hor}}(E=1)=4T_{H}$ was obtained previously in \cite{Barbado2011}.

\bibliographystyle{utcaps}
\providecommand{\href}[2]{#2}\begingroup\raggedright\endgroup
\end{document}